\documentclass[a4paper,aps,prd,twocolumn,10pt,superscriptaddress,nofootinbib,showpacs]{revtex4}
\usepackage{graphicx,amsmath,amsfonts,amssymb,revsymb,dcolumn,epsfig,bm,hyperref,xspace}
\usepackage[latin1]{inputenc}
\usepackage{color}

\newcommand{\dd}{\mathrm{d}}
\newcommand{\lie}{\pounds}
\newcommand{\del}{\partial}

\newcommand{\Poisson}[2]{\left\{#1,\;#2\right\}}
\DeclareFontFamily{U}{mathx}{\hyphenchar\font45}
\DeclareFontShape{U}{mathx}{m}{n}{<-> mathx10}{}
\DeclareSymbolFont{mathx}{U}{mathx}{m}{n}
\DeclareMathAccent{\widebar}{0}{mathx}{"73}
\newcommand{\backdec}[1]{\bar{#1}}
\newcommand{\wbackdec}[1]{\widebar{#1}}
\newcommand{\g}{g}
\newcommand{\dg}{{\delta\g}}
\newcommand{\bg}{{\backdec{g}}}
\newcommand{\n}{v}
\newcommand{\bn}{{\backdec{\n}}}

\newcommand{\h}{\gamma}
\newcommand{\bh}{\backdec{\h}}

\newcommand{\hp}[1]{\h\left[#1\right]}
\newcommand{\bhp}[1]{\bh\left[#1\right]}
\newcommand{\cd}{\nabla}
\newcommand{\bcd}{\wbackdec{\nabla}}
\newcommand{\scd}{D}
\newcommand{\bscd}{\wbackdec{\scd}}
\newcommand{\scp}{\parallel}
\newcommand{\blap}{\wbackdec{\scd}^2}
\newcommand{\blapK}{\blap_{\K}}

\newcommand{\R}{R}

\newcommand{\bR}{\wbackdec{\R}}
\newcommand{\G}{G}
\newcommand{\bG}{\wbackdec{\G}}
\newcommand{\kp}{\kappa}

\newcommand{\ddg}{\xi}

\newcommand{\LL}{{\mathcal F}}
\newcommand{\LLa}{\LL_{a}{}}
\newcommand{\LLb}{\LL_{b}{}}

\newcommand{\A}{\phi}
\newcommand{\B}{B}
\newcommand{\C}{C}
\newcommand{\CS}{\psi}
\newcommand{\CSI}{\Psi}
\newcommand{\CSvms}{\mathcal{U}}
\newcommand{\MS}{\zeta}
\newcommand{\CST}{\CS^{\mathsf{t}}}
\newcommand{\CTD}{W}
\newcommand{\BS}{\mathcal{\B}}
\newcommand{\BV}{\mathtt{\B}}
\newcommand{\CSD}{\mathcal{E}}
\newcommand{\CV}{\mathtt{F}}
\newcommand{\zz}{\mathtt{z}}
\newcommand{\AQT}[1]{Q^{\text{(#1)}}{}}
\newcommand{\EC}{\mathcal{K}}

\newcommand{\EX}{\Theta}
\newcommand{\dEX}{{\delta\EX}}

\newcommand{\bEX}{\wbackdec{\EX}}
\newcommand{\SH}{\sigma}
\newcommand{\dSH}{\delta\sigma}
\newcommand{\dSHs}{\dSH^{\text{s}}{}}
\newcommand{\dSHv}{\dSH^{\text{v}}{}}

\newcommand{\SR}{\mathcal{R}}
\newcommand{\dSR}{{\delta\SR}}
\newcommand{\bSR}{\wbackdec{\SR}}
\newcommand{\K}{K}
\newcommand{\bK}{\wbackdec{\K}}
\newcommand{\ac}{a}

\newcommand{\T}{T}

\newcommand{\bT}{\wbackdec{T}}

\newcommand{\mf}{\varphi}
\newcommand{\dmf}{\delta\mf}

\newcommand{\vms}{\mathcal{V}}
\newcommand{\bmf}{\backdec{\mf}}

\newcommand{\p}{p}
\newcommand{\dpp}{\delta\p}

\newcommand{\ed}{\rho}

\newcommand{\ded}{\delta\ed}

\newcommand{\bp}{\backdec{p}}
\newcommand{\bed}{\backdec{\ed}}

\newcommand{\LA}{\mathcal{L}}

\newcommand{\AC}{S}
\newcommand{\bLA}{\wbackdec{\LA}}
\newcommand{\bAC}{\wbackdec{\AC}}

\newcommand{\dLA}{{\delta\LA}}

\newcommand{\pot}{\mathit{U}}
\newcommand{\bpot}{\wbackdec{U}}
\newcommand{\kin}{\text{k}}
\newcommand{\mat}{\text{m}}

\newcommand{\HA}{\mathcal{H}}
\newcommand{\dHA}{{\delta\HA}}
\newcommand{\pCSD}{\Pi_{\CSD}}
\newcommand{\pCST}{\Pi_{\CST}}
\newcommand{\pvms}{\Pi_{\vms}}

\newcommand{\pMS}{\Pi_\MS}
\newcommand{\pa}{\Pi_a}
\newcommand{\pbmf}{\Pi_{\bmf}}

\newcommand{\BGE}{E}
\newcommand{\cV}{\widetilde{V}}
\newcommand{\cK}{\widetilde{\K}}
\newcommand{\vu}{\mathrm{V}}
\newcommand{\pct}{\del_\text{ct}}
\begin{document}
\title{Scalar Field Perturbations with Arbitrary Potentials in Quantum Backgrounds}

\author{F.~T.~Falciano} \email{ftovar@cbpf.br}
\author{Nelson Pinto-Neto} \email{nelson.pinto@pq.cnpq.br}
\author{Sandro Dias Pinto Vitenti} \email{vitenti@cbpf.br}

\affiliation{Centro Brasileiro de Pesquisas F\'{\i}sicas,
Rua Dr.\ Xavier Sigaud 150 \\
22290-180, Rio de Janeiro -- RJ, Brasil}

\date{\today}

\begin{abstract}

In this paper it is shown how to obtain,
without ever using the background classical equations of motion, a simple second order Hamiltonian 
involving the Mukhanov-Sasaki variable describing 
quantum linear scalar perturbations for the case of scalar
fields with arbitrary potentials and arbitrary spacelike hyper-surfaces. It is a
generalization of previous works, where the scalar field potential was
absent and the spacelike hypersurfaces were flat. This was possible
due to the implementation of a new method, together with the Faddeev-Jackiw 
procedure for the constraint reduction. The resulting Hamiltonian can then 
be used to study the evolution of quantum cosmological perturbations in quantum
backgrounds.

\end{abstract}

\pacs{98.80.Es, 98.80.-k, 98.80.Jk}

\maketitle

\section{Introduction}

The usual theory of cosmological perturbations, with their simple equations, relies essentially on the assumptions that the background is described by pure classical General Relativity, while the perturbations thereof stem from quantum fluctuations. It is a semiclassical approach, where the background is classical and the perturbations are quantized, and the fact that the background satisfies Einstein's equations is heavily used in the simplification of the equations.

The next and more fundamental conceptual step is to consider the more general situation where quantum effects are present already on the background geometry. In this regime, the usual semi-classical treatment of cosmological perturbations is no longer valid. Even though quantizing simultaneously the homogeneous background and their linear perturbations is still far from the full theory of quantum gravity, one can consider the inclusion of quantum effects in the dynamics of the background homogeneous model as an important improvement to the usual semi-classical approach~\cite{Halliwell1985}. Note, however, that this program prevent us from using the classical background equations, as it is usually done, to turn the full second order action into a simple treatable system. 

Notwithstanding, it has already been shown that it is possible to simplify the Einstein-Hilbert action through canonical transformation techniques for a barotropic perfect fluid and scalar fields without potential in a flat spatial section Friedmann model~\cite{Peter2005,Pinho2007,Falciano2009}. In these frameworks, the Hamiltonian constraint of General Relativity up to second order was put in the form ${\cal{H}} = {\cal{H}}^{(0)} + {\cal{H}}^{(2)}$, where ${\cal{H}}^{(0)}$ is the background Hamiltonian constraint while ${\cal{H}}^{(2)}$ is the Hamiltonian constraint for the perturbations. The natural and more general way to Dirac quantize the theory is to impose the annihilation of the wave functional by the full Hamiltonian constraint ${\cal{H}}$, $\hat{{\cal{H}}}|\Psi> = 0$, which imposes a quantization of the background and perturbations. Due to the simplifications obtained, it was possible to solve the quantum equations for the background and perturbations in many circumstances, and calculate their observational consequences. 

The scenarios obtained describe cosmological perturbations of quantum mechanical origin evolving in a non-singular homogeneous and isotropic background, in which quantum effects replace the usual classical singularity by a bounce. The physical properties of these cosmological models were analyzed in many papers \cite{Peter2006,Peter2007,Peter2008,Falciano2007,Pinto-Neto2005,AcaciodeBarros1998,Alvarenga2002,Pinto-Neto2003}, and they proved to be complementary or even competitive with usual inflationary models as long as they are capable to lead to almost scale invariant spectra of long-wavelength cosmological perturbations.

The aim of this paper is to improve the previous formalism and to extend the known results to a scalar field with arbitrary potential in a Friedmann background with arbitrary spacelike hyper-surfaces. In order to carry out this work, we use the same techniques of Ref.~\cite{Vitenti2012a}: we implement a set of variable transformations along with the Faddeev-Jackiw~\cite{Faddeev1988,Jackiw1993} reduction method, rather than the Dirac formalism. The resulting action and Hamiltonian up to second order then become very simple and suitable for canonical quantization.

Besides the motivation related to the quantization procedure, our choice of variables used to write down the second order Lagrangian simplifies significantly the calculations involved. This simplification allows us to obtain all expressions without choosing a gauge. This is an important advantage since we have shown in Ref.~\cite{Vitenti2012} that the choice of a gauge implies an additional assumption that the perturbations should remain small in this gauge.

The paper is organized as follows: in the next section we define some relevant geometrical objects and settle down the notation and conventions. In Section III we review the methods to obtain the second order gravitational Lagrangian for geometrical perturbations around a homogeneous and isotropic geometry with arbitrary spacelike hyper-surfaces, while in Section IV we obtain the second order matter Lagrangian for a canonical scalar field with arbitrary potential. All the results are obtained without assuming the validity of the background Einstein's equations. In Section V we combine the results of Sections III and IV in order to obtain the full simplified general relativistic action up to second order terms, and its resulting full simplified Hamiltonian constraint in the desired form
${\cal{H}} = {\cal{H}}^{(0)} + {\cal{H}}^{(2)}$, ready to be Dirac quantized. We end up with the conclusions. 

\section{Geometry and Spacetime Foliation}\label{sec:three_one}

The present paper follows closely the definitions and terminologies used in \cite{Vitenti2012a} but, for sake of completeness, we shall briefly define some relevant geometrical objects and fix our notation. 

The spacetime Lorentzian metric $\g_{\mu\nu}$ has signature $(-1,1,1,1)$ and the covariant derivative compatible with this metric is represented by $\cd_\mu$, i.e., $\cd_\alpha \g_{\mu\nu} = 0$. 

We define the foliation of the spacetime through a normalized timelike vector field $\n^\mu$ normal to each spatial section ($\n^\mu\n_\mu = -1$). The foliation induces a metric in the hyper-surfaces as $\h_{\mu\nu} = \g_{\mu\nu} + \n_\mu\n_\nu$ which projects non-spatial vectors into the hyper-surfaces. 

For an arbitrary tensor $M_{\phantom a \phantom a \mu_1\dots\mu_m}^{\nu_1\dots\nu_k}$ the projector is defined as
\begin{equation*}
\hp{M^{\nu_1\dots\nu_k}_{\phantom a \phantom a \mu_1\dots\mu_m}} \equiv \h_{\mu_1}^{\phantom a \alpha_1}\dots\h_{\mu_m}^{\phantom a\alpha_m}\h_{\phantom a \beta_1}^{\nu_1}\dots\h_{\phantom a \beta_k}^{\nu_k}M^{\beta_1\dots\beta_k}_{\phantom a \phantom a \alpha_1\dots\alpha_m},
\end{equation*}
and we shall call a spatial object any tensor that is invariant under this projection, i.e. $\hp{M_{\phantom a \phantom a \mu_1\dots\mu_m}^{\nu_1\dots\nu_k}} = M_{\phantom a \phantom a \mu_1\dots\mu_m}^{\nu_1\dots\nu_k}$. 

The covariant derivative compatible with the spatial metric $\h_{\mu\nu}$ is
\begin{equation}
\scd_\alpha M_{\mu_1\dots\mu_m}{}^{\nu_1\dots\nu_k} = \hp{\cd_\alpha M_{\mu_1\dots\mu_m}{}^{\nu_1\dots\nu_k}}.
\end{equation}
from which we can define the spatial Riemann curvature tensor as
\begin{equation}\label{eq:def:SR}
\SR_{\mu\nu\alpha}{}^\beta A_\beta \equiv [\scd_\mu\scd_\nu - \scd_\nu\scd_\mu]A_\alpha,
\end{equation}
with $A_\alpha$ an arbitrary vector field. The spatial Laplacian is represented by the symbol $\scd^2$, i.e., $\scd^2 \equiv \scd_\mu\scd^\mu$, and we denote the contraction with the normal vector field $\n^\mu$ with a index $\n$, e.g., $M_{\alpha \n} \equiv M_{\alpha \beta}\n^\beta$.

The derivative of the velocity field defining the foliation can be decomposed as
\begin{equation}\label{eq:def:EC:2}
\cd_\mu\n_\nu = \EC_{\mu\nu} - \n_\mu\ac_\nu,
\end{equation}
with the acceleration and the extrinsic curvature defined respectively as $\ac_\mu \equiv \n^\gamma\cd_\gamma\n_\mu$ and $\EC_{\mu\nu} \equiv \hp{\cd_\mu\n_\nu}$. In addition, from the extrinsic curvature we define the expansion factor and the shear \footnote{The Frobenius theorem guaranties that for a global spatial sectioning the normal field satisfy $\n_{[\alpha}\cd_{\mu}\n_{\nu]} = 0$, which can be expressed as $\cd_{[\mu}\n_{\nu]} = \ac_{[\mu}\n_{\nu]}$. Therefore, for a global spatial sectioning the vorticity is null, i.e., $\EC_{[\mu\nu]} = 0$.} as
\begin{equation}
\EX \equiv \EC_\mu{}^\mu, \qquad \SH_{\mu\nu} \equiv \EC_{\mu\nu} - \frac{\EX}{3}\h_{\mu\nu}.
\end{equation}

In what follows we are going to study perturbations of the metric tensor, hence, we are impel to distinguish between the actual physical metric $\g_{\mu\nu}$ and a given fiducial background metric $\bg_{\mu\nu}$ that can be used as reference. We shall also assume that the physical metric $\g_{\mu\nu}$ can be seen as close to the fiducial metric in the sense that their difference $\dg_{\mu\nu} = \g_{\mu\nu} - \bg_{\mu\nu}$ can be treated perturbatively (see \cite{Vitenti2012} for details). 

The background and the perturbed tensors shall have their indices always raised and lowered by the background metric. Therefore, we must distinguish between the object formed by raising the indices of $\dg_{\mu\nu}$ and the difference between the inverse metric and its background value i.e. $\dg^{\mu\nu} \equiv \g^{\mu\nu} - \bg^{\mu\nu}$. To avoid any possible confusion, we define the tensor $\ddg_{\mu\nu}$ and its covariant form as
\begin{equation}\label{eq:def:ddg}
\ddg_{\mu\nu} \equiv \g_{\mu\nu} - \bg_{\mu\nu},\quad \ddg^{\mu\nu}\equiv\bg^{\mu \alpha}\bg^{\nu \beta}\ddg_{\alpha \beta},
\end{equation}
which will always describe the perturbations of the metric tensor.
 
The covariant derivative compatible with the background metric is represented by the symbol $\bcd$ or by a semicolon ``$\, ;\, $'', i.e. $\bg_{\mu\nu;\gamma} \equiv \bcd_\gamma\bg_{\mu\nu} = 0$. 

Using a background foliation described by the normal vector field $\bn^\mu$, we can again define the background projector $\bh_{\mu\nu}$, spatial derivative $\bscd_\mu$ and the background spatial Riemann tensor $\bSR_{\mu\nu\alpha}{}^\beta$ in the same way we have done for objects derived from the foliation of the physical manifold.

We use the symbol ``$ {}_{\scp} $'' to represent the background spatial derivative, i.e., $T_{\scp\mu} \equiv \bscd_\mu T$ for any tensor $T$. Finally, we define the dot operator of an arbitrary tensor as
\begin{equation}
\dot{M}_{\mu_1\dots\mu_m}{}^{\nu_1\dots\nu_k} \equiv \bhp{\lie_{\bn}{}M_{\mu_1\dots\mu_m}{}^{\nu_1\dots\nu_k}}.
\end{equation}

\section{Gravitational Action}

In order to have a linear dynamical system, we must expand its associated action functional at least to second order in the perturbations. In \cite{Vitenti2012a}, we have performed the simplification of the second order action considering a system composed of gravity and a generic perfect fluid as its matter content.

The full simplification can be completed only once we consider simultaneously the gravitational and the matter sector. Notwithstanding, the coupling between matter and gravity occurs only in the matter Lagrangian. As a result, a great amount of work in the perturbative expansion of the gravitational sector can be done independently from the perturbative expansion of the matter fields. 

Accordingly, much of the work in the expansion of the gravitational sector has already been done in \cite{Vitenti2012a}. Here we shall reproduce only its essential steps without lack of clarity for the reader inasmuch as the same technique will be used in detail for the scalar field matter Lagrangian in section \ref{sec:scalarfield} .

As it is well known, the pure gravitational part of the action reads
\begin{equation}\label{eq:def:ACg}
\AC_\g = \int\dd^4x\LA_\g, \qquad \LA_\g \equiv \frac{\sqrt{-\g}\R}{2\kp},
\end{equation}
where $\kp = 8\pi{}G/c^4$, $G$ is the gravitational constant, $c$ the speed of light and $\R$ is the curvature scalar. The expansion in the curvature tensor induced by Eq.~\eqref{eq:def:ddg} can be described by a tensor $\LL_{\mu\nu}{}^\alpha$ that is defined as the difference between the covariant derivative of the perturbed and of the background metrics, i.e.
\begin{align}
&(\cd_\mu-\bcd_\mu)A_\nu = \LL_{\mu\nu}{}^\beta A_\beta ,\nonumber\\
&\LL_{\alpha\beta}{}^\gamma = -\frac{\g^{\gamma\sigma}}{2}\left(\g_{\sigma\beta;\alpha} + \g_{\sigma\alpha;\beta} - \g_{\alpha\beta;\sigma}\right). \label{eq:LL}
\end{align}
Let us re-emphasize that a semicolon represents covariant derivative with respect to the background metric. Thus, even though we have a Riemannian physical manifold, in general $\g_{\alpha\beta;\sigma}\neq0$. Making use of this tensor we can construct two covectors as
\begin{align}
\LLa_\alpha &\equiv \LL_{\nu\alpha}{}^\nu = - \frac{\ddg_{;\alpha}}{2}, \\
\LLb_\mu &\equiv \bg^{\alpha\beta}\LL_{\alpha\beta\mu} = -\ddg_\mu{}^{\sigma}{}_{;\sigma} + \frac{\ddg_{;\mu}}{2}.
\end{align}

The physical Riemann tensor associated with $\g_{\mu\nu}$ can be written in terms of the background Riemann tensor $\bR_{\mu\nu\alpha}{}^\beta$ related to $\bg_{\alpha\beta}$ as
\begin{equation}\label{eq:pR:R}
\R_{\mu\nu\alpha}{}^\beta = \bR_{\mu\nu\alpha}{}^\beta + 2\LL_{\alpha[\nu}{}^\beta{}_{;\mu]} + 2\LL_{\alpha[\mu}{}^\gamma\LL_{\nu]\gamma}{}^\beta.
\end{equation}
Thus, the expansion of the curvature scalar up to second order reads
\begin{equation}\label{eq:pSR:SR}
\begin{split}
\R &\approx \bR + \bR_{\mu\alpha}\dg^{\mu\alpha} + (\LLa^\mu - \LLb^\mu)_{;\mu} + \LLb^\mu\LLa_\mu \\
&- \LL_{\mu\nu\alpha}\LL^{\mu\alpha\nu} + (\LLa_{\nu;\mu}-\LL_{\mu\nu}{}^\gamma{}_{;\gamma})\dg^{\mu\nu},
\end{split}
\end{equation}
with $\bR$ and $\bR_{\mu\alpha}$ being respectively the scalar curvature and the Ricci tensor of the background.

\subsection{Second Order Gravitational Lagrangian}

The above expansion Eq.~\eqref{eq:pSR:SR} shows that with respect to the tensor $\LL_{\mu\nu}{}^\alpha$ the second order Lagrangian assumes a simple and compact form. Indeed, ignoring a surface term, the gravitational Lagrangian decomposes in three terms as
\begin{equation}
\LA_\g = \bLA_\g + \dLA_\g^{(1)} + \dLA_\g^{(2)} .
\end{equation}
The background and the first order terms are given by
\begin{align}
\bLA_\g = \sqrt{-\bg}\frac{\bR}{2\kp}, \qquad \dLA_\g^{(1)}=-\frac{\sqrt{-\bg}}{2\kp}\bG_{\mu\nu}\ddg^{\mu\nu},
\end{align}
with $\bG_{\mu\nu}$ being the background Einstein tensor. The second order Lagrangian can be split in a kinetic term $\dLA_{\g\kin}^{(2)}$ that includes derivatives of the perturbations and a potential term $\dLA_{\g\pot}^{(2)}$ without their derivatives. These terms read
\begin{align}
\dLA_{\g\kin}^{(2)} &= \frac{\sqrt{-\bg}}{2\kp}\left[\LL^{\mu\nu\gamma}\LL_{\gamma(\mu\nu)} - \LLa_\mu\LLb^\mu\right],\label{eq:def:dLAk2}\\
\dLA_{\g\pot}^{(2)} &= \frac{\sqrt{-\bg}}{2\kp}\left[\bG_{\mu\nu}+\frac{\bg_{\mu\nu}}{4}\bR\right]\ddg^{\mu}{}_\alpha\left[\ddg^{\alpha\nu}-\frac{\bg^{\alpha\nu}\ddg}{2} \right].\label{eq:def:dLAp2}
\end{align}

Regardless of the compact and elegant form of the above expression, the perturbed degrees of freedom are in fact encoded in the perturbations of metric tensor. Thus, we need to express the $\LL_{\mu\nu}{}^\gamma$ tensor in terms of the perturbed kinematic parameters associated with the spatial slicing. For a given background foliation $\bn^\mu$, the metric perturbation can be decomposed as
\begin{equation}\label{eq:ddg:decomp}
\ddg_{\mu\nu} = 2\phi\bn_\mu\bn_\nu+2\B_{(\mu}\bn_{\nu)}+2\C_{\mu\nu},
\end{equation}
where by construction
\begin{align*}
&\phi\equiv\frac12 \ddg_{\bn\bn}, &\B^\mu \equiv \bhp{\ddg_{\bn}{}^{\mu}},\\
&\C_{\mu\nu} \equiv \frac{\bhp{\ddg_{\mu\nu}}}{2}, &\C \equiv \frac{\ddg_{\mu\nu}\bh^{\mu\nu}}{2}.
\end{align*}

Besides, we can continue the decomposition in terms of their tensorial nature. By taking into account the decomposition in terms of scalar, vector and tensor objects (see~\cite{Stewart1990}) we are able to write them as
\begin{align*}
\B_\mu &= \BS_{\scp\mu} + \BV_\mu, \\
\C_{\mu\nu} &= \CS\h_{\mu\nu} - \CSD_{\scp\mu\nu} + \CV_{(\nu\scp\mu)} + \CTD_{\mu\nu},
\end{align*}
which should be traceless $ \CTD_\mu{}^\mu = 0$ and divergence free $\BV^\mu{}_{\scp\mu} = \CV^\mu{}_{\scp\mu} = \CTD_\mu{}^\nu{}_{\scp\nu}=0$. In terms of this decomposition, the perturbation in the expansion factor translate into
\begin{equation}\label{eq:FLRW:dEX}
\dEX = \blap\dSHs  + \bEX\A + 3\dot{\CS},
\end{equation}
and the perturbed shear tensor
\begin{equation}\label{eq:def:FLRW:sCI:decomp}
\dSH_{\mu\nu} = \left(\bscd_{(\mu}\bscd_{\nu)}-\frac{\bh_{\mu\nu}\blap}{3}\right)\dSHs + \dSHv_{(\nu\scp\mu)} + \dot{\CTD}_\mu{}^\alpha\bh_{\alpha\nu},
\end{equation}
where we have defined
\begin{equation}\label{eq:FLRW:dSHs}
\dSHs \equiv \left(\BS-\dot{\CSD} + \frac{2}{3}\bEX\CSD\right), \quad \dSHv^{\alpha} \equiv \BV^\alpha + \dot{\CV}^\alpha.
\end{equation}
The perturbed spatial Ricci tensor for this foliation is
\begin{align*}
\bhp{\dSR_\mu{}^\bn} &= 0, \\
\bhp{\dSR_\bn{}^\nu} &= -2\bK[\BS^{\scp\nu} + \BV^\nu], \\
\bhp{\dSR_\mu{}^\nu} &= -\CS_{\scp\mu}{}^{\scp\nu} - \bh_\mu{}^\nu[\blap + 4\bK]\CS - [\blap - 2\bK]\CTD_\mu{}^\nu, \\
\dSR &= -4\blapK\CS,
\end{align*}
where we have defined the operator $\blapK \equiv \blap + 3\bK$. The detailed calculation of these quantities for a general background and for a Friedmann background can be found at Appendix~A and Section~IIIA of Ref.~\cite{Vitenti2012a}, respectively. There the calculations were done for an arbitrary background and without fixing the spatial hypersurfaces, hence, the slicing of the physical manifold need not be the same as the background spatial hypersurfaces. The only assumption was that both spatial sectioning were global and the background foliation is geodesic, i.e., $\bn^\mu\bcd_\mu\bn^\nu = 0$. However, when applying for a Friedmann background we considered the same foliation for the kinematic variables where the fields $\n_\mu$ and $\bn_\mu$ differ only by their normalization\footnote{For more details see also the discussion in the beginning of Appendix~B of~\cite{Vitenti2012a}}.

As we have shown in \cite{Vitenti2012a}, one can expand the gravitational action up to second order without the need to impose any symmetry for the background foliation. The full procedure is however very involved. In contrast, if we assume that the background is a Friedmann-Lema\^{\i}tre-Robertson-Walker (FLRW) universe, the gravitational second order Lagrangian in terms of these quantities reads $$\dLA_\g^{(2)}=\dLA_\g^{(2,s)}+\dLA_\g^{(2,v)}+\dLA_\g^{(2,t)},$$ where each represents one of the independent sectors, namely, the scalar, vector and tensorial. Explicitly, they are respectively
\begin{align}
\frac{\dLA_\g^{(2,s)}}{\sqrt{-\bg}} &= \frac{1}{3\kp}\left(\blap\dSHs\blapK\dSHs-\dEX^2\right) \nonumber \\ \label{eq:FLRW:dLA2:s}
&+ \left(\frac{\CS}{2}-\A\right)\frac{\dSR}{2\kp} + \frac{\bG_{\bn\bn}}{2\kp}(\B_\gamma\B^\gamma - \phi^2 - 2\C\phi) \nonumber\\
&+ \frac{\bG_{\mu \nu}\bg^{\mu \nu}}{6\kp} (2\C_\gamma{}^\nu\C_\nu{}^\gamma-\C^2), \\
\frac{\dLA_\g^{(2,v)}}{\sqrt{-\bg}} &= \frac{\dSHv_{(\alpha\scp\nu)}\dSHv^{(\alpha\scp\nu)}}{2\kp}, \label{eq:FLRW:dLA2:v}\\ 
\frac{\dLA_\g^{(2,t)}}{\sqrt{-\bg}} &= \frac{\dot{\CTD}_\nu{}^\gamma\dot{\CTD}_\gamma{}^\nu  + \CTD_\mu{}^\nu(\blap - 2K)\CTD_\nu{}^\mu}{2\kp}. \label{eq:FLRW:dLA2:t}
\end{align}

\section{Scalar Field Action} \label{sec:scalarfield}

In close analogy with the perturbation procedure develop for the gravitational sector, we shall now perturb the matter Lagrangian up to second order. We will consider the case where the matter content is described by a real scalar field with an arbitrary algebraic potential $\pot(\mf)$. The action for a scalar field with potential reads
\begin{equation}
\AC_\mat = \int\dd^4 x\sqrt{-\g}\left(\frac{\cd_\alpha\mf\cd^\alpha\mf}{2} + \pot(\mf)\right).
\end{equation}

It is straightforward to show that its energy momentum tensor can be written as
\begin{equation}\label{eq:def:S:Tab}
\begin{split}
\T_{\mu\nu} &\equiv \frac{-2}{\sqrt{-\g}}\frac{\delta\AC_m}{\delta\g^{\mu\nu}}, \\ 
&= \cd_\mu\mf\cd_\nu\mf - \g_{\mu\nu}\left(\frac{\cd_\alpha\mf\cd^\alpha\mf}{2} + \pot(\mf)\right).
\end{split}
\end{equation}

As a matter of consistency, the energy-momentum tensor must be compatible with the symmetries of the spacetime metric. Thus, by considering the FLRW universe as the background structure, it follows that the given projection of the energy-momentum tensor must be zero
\begin{equation}
\hp{\bT_\mu{}^\n} = \dot{\bmf}\bscd_\mu\bmf = 0,
\end{equation}
where $\bmf$ is the background scalar field, such that, $\dmf \equiv \mf - \bmf$ defines the perturbation on $\bmf$.
This restriction implies that $\bscd_\mu\bmf = 0$. In this manner, we can rewrite the derivative of the background field as $\bcd_\mu\bmf = -\bn_\mu\dot{\bmf}$. The background energy momentum tensor is then
\begin{equation}
\bT_{\mu\nu} = \bed\bn_\mu\bn_\nu + \bp\h_{\mu\nu},\quad \bed \equiv \frac{\dot{\bmf}^2}{2} + \bpot,\quad \bp \equiv \frac{\dot{\bmf}^2}{2} - \bpot,
\end{equation}
where $\bpot \equiv \pot(\bmf)$. An expansion of the components of the energy-momentum tensor shows that its perturbed quantities are given by
\begin{align}
\ded &= \dot{\dmf}\dot{\bmf} + \phi\dot{\bmf}^2 + \pot_{\bmf}\dmf, \label{eq:def:ded}\\ 
\dpp &= \dot{\dmf}\dot{\bmf} + \phi\dot{\bmf}^2 - \pot_{\bmf}\dmf, \label{eq:def:dpp}\\
\vu_\mu &= -\frac{\bscd_\mu\dmf}{\dot{\bmf}},\quad\delta\Pi_\mu{}^\nu = 0,\label{eq:def:vu}
\end{align}
where $\pot_{\bmf}$ means $\pot_{\bmf} =\frac{ \del{}\pot}{\del\mf}\Big|_{\bmf}$. Using a power expansion in the perturbed variables, we can separate the matter action order by order in the perturbations,
\[
\AC_{\mat} =\bAC_{\mat} +\AC_{\mat}^{(1)} +\AC_{\mat}^{(2)},
\]
with
\begin{align}
\bAC_{\mat}&= \int\dd^4 x\sqrt{-\bg}\left(\frac{\bcd_\alpha\bmf\bcd^\alpha\bmf}{2} + \bpot\right), \label{eq:def:bAC}\\
\AC_{\mat}^{(1)}&= \int\dd^4{}y\left(\wbackdec{\frac{\delta \AC}{\delta\mf}}\dmf
+\frac{\sqrt{-\bg}}{2}\bT^{\mu \nu}\ddg_{\mu\nu}\right),\qquad \label{eq:def:ACm1}\\ \label{eq:def:ACm2}
\AC_{\mat}^{(2)} &= \int\dd^4{}x\dLA_\mat^{(2)},
\end{align}
where the over line in the variations means that these expression are evaluated at the background fields, i.e., ($\bmf$, $\bg_{\mu\nu}$).

A direct calculation shows that the second order Lagrangian $\dLA_\mat^{(2)}$ can be decomposed in four terms $\dLA_\mat^{(2)}=\dLA_\mat^{(2,1)}+ \dLA_\mat^{(2,2)} + \dLA_\mat^{(2,3)} + \dLA_\mat^{(2,4)}$, which are respectively,
\begin{align}
\dLA_\mat^{(2,1)} &= \int\dd^4{}y\sqrt{-\bg(y)}\wbackdec{\frac{\delta \T^{\mu\nu}(y)}{\delta\mf(x)}}\frac{\dmf(x)\ddg_{\mu\nu}(y)}{2}, \\
\dLA_\mat^{(2,2)} &= \int\dd^4{}y\sqrt{-\bg(y)}\wbackdec{\frac{\delta\T^{\alpha\beta}(y)}{\dg_{\mu\nu}(x)}}\frac{\ddg_{\mu\nu}(x)\ddg_{\alpha\beta}(y)}{4}, \\
\dLA_\mat^{(2,3)} &= \frac18\sqrt{-\bg(x)}\ \bT^{\alpha\beta}(x)\ddg(x)\ddg_{\alpha\beta}(x), \\
\dLA_\mat^{(2,4)} &= \int\dd^4{}y\wbackdec{\frac{\delta^2\AC_m}{\delta\mf(x)\delta\mf(y)}}\frac{\dmf(x)\dmf(y)}{2},
\end{align}
with $\ddg \equiv \ddg_{\mu\nu}\bg^{\mu\nu}$. 

Combining all these contributions, the second order matter Lagrangian reads
\begin{equation}\label{eq:def:dLAm2}
\begin{split}
\dLA_\mat^{(2)} &= l_{g\mf}+l_{\mf}-\sqrt{-\bg}\left(\B_\mu\B^\mu -\phi^2- 2\C\phi\right)\frac{\bed}{2}\\
&- \sqrt{-\bg}\left[2\C_\mu{}^\nu\C^\mu{}_\nu-\C^2\right]\frac{\bp}{2},
\end{split}
\end{equation}
where we have combined in $l_{g\mf}$ terms that mix metric perturbations with $\dmf$ and in the $l_{\mf}$ terms that are simply quadratic in the latter, i.e.
\begin{align}\label{eq:def:lgmf}
\frac{l_{g\mf}}{\sqrt{-\bg}}&\equiv\phi\ded + \C\dpp - \left(\B_\mu\vu^\mu + \frac{\phi^2}{2} + \C\phi\right)\dot{\bmf}^2, \\ \label{eq:def:lmf}
\frac{l_{\mf}}{\sqrt{-\bg}} &\equiv \frac{1}{2}\left(\dot{\dmf}^2-\bscd_\mu\dmf\bscd^\mu\dmf-\pot_{\bmf\bmf}\dmf^2\right).
\end{align}
Note that the specific combination of perturbed fields of the last two parts of Eq.~\eqref{eq:def:dLAm2} also appears in Eq.~\eqref{eq:FLRW:dLA2:s}. When we combine them in the full second order Lagrangian, their sum forge terms that are proportional to the background dynamical equations, namely the time-time and the space-space Einstein's equations. 

These particular expressions and the background equations of motion will repeatedly appear in our subsequent calculations. Therefore, it is convenient to define the following quantities
\begin{align*}
\BGE_{\bmf} &\equiv \ddot{\bmf}+\dot{\bmf}\bEX+\pot_{\bmf},\\
\BGE_\bn &\equiv \bG_{\bn\bn} - \kp\bed,\\
\BGE_{\bh} &\equiv \frac{\bG_{\mu\nu}\bh^{\mu\nu}}{3} - \kp\bp,\\
\BGE_{\bg} &\equiv \BGE_\bn + \BGE_{\bh} = \dot{\bEX} + \frac{3\kp\dot{\bmf}^2}{2}-3\bK.
\end{align*}

The suitable calculation needed to simplify the perturbed action contains a lot of terms and can become unmanageable. Thus, every change of variable that simplifies the expressions are in fact crucial. In this spirit, we shall make two changes of variable that seems devoided of physical significance. They should be viewed only as an intermediary step that simplifies the equations. Thus, we define $$\vms \equiv -\frac{\dmf}{\dot{\bmf}},\quad \CST \equiv \frac{\C}{3} = \CS - \frac13\blap\CSD.$$

Using the variables defined above and the others defined in Eqs.~\eqref{eq:FLRW:dEX}-\eqref{eq:FLRW:dSHs} and \eqref{eq:def:ded}-\eqref{eq:def:vu}, the expression for $l_{g\mf}$ changes to 
\begin{equation}\label{eq:def:lgmf2}
\frac{l_{g\mf}^*}{\sqrt{-\bg}} = \dEX\dot{\bmf}^2\vms + 3\CST\vms\dot{\bmf}\BGE_{\bmf},
\end{equation}
where we discarded the surface term $\pct(3\CST\sqrt{-\bg}\dot{\bmf}\dmf)$. Applying the same reasoning to $l_{\mf}$ we obtain
\begin{equation}\label{eq:def:lmf2}
\begin{split}
&\frac{l_{\mf}^*}{\sqrt{-\bg}} = \frac{\left(\ded-\bEX\dot{\bmf}^2\vms\right)^2}{2} + \frac{\vms^2}{2}\left(\dot{\bmf}^2\BGE_{\bg}-\pot_{\bmf}\BGE_{\bmf}\right)+\\
&+ \frac{\dot{\bmf}^2}{2}\left(\vms\blapK\vms-\frac{3\kp\dot{\bmf}^2\vms^2}{2}\right),
\end{split}
\end{equation}
where we discarded the term $\pct(\sqrt{-\bg}\dmf^2(\pot_{\bmf}/\dot{\bmf}+\bEX)/2)$. The symbol ${}^*$ in the expressions above means that we have removed from $l_{\mf}$ and $l_{g\mf}$ the terms which cancel out in the sum, i.e., $l_{\mf} + l_{g\mf} = l_{\mf}^* + l_{g\mf}^*$.

The terms proportional to $\BGE_{\bmf}$ and $\BGE_{\bg}$ in the above equations can be eliminated without assuming the validity of the background equations of motion. As discussed in \cite{Vitenti2012a}, it is legitimate to redefine our basic perturbed variables by adding second order terms to them. 

In this way, we redefine for instance the variable $\A$ into a new variable $\A_\text{new}$ in such a way that $\A_\text{new} = \A + \delta{}f$ with $\delta{}f$ being a certain combination of second order perturbations. Note that, by construction, both variables agree at first order and are different only at second order. 

Thus, this kind of transformation changes only the second order part of the Lagrangian by adding a new term coming from the first order part in the old variables. The fact that it comes from the first order Lagrangian makes them proportional to the background equation of motion
\begin{equation*}
\frac{2\kp\dLA^{(1)}(\A)}{\sqrt{-\bg}} \approx \frac{2\kp\dLA^{(1)}(\A_\text{new})}{\sqrt{-\bg}} + 2\BGE_{\bn}\delta{}f.
\end{equation*}

We shall profit from this freedom in defining the perturbed variables and cancel the terms proportional to $\BGE_{\bmf}$ by redefining the $\vms$ variable as
\begin{align}\label{eq:vmstrans}
\vms &\rightarrow \vms +\frac12\pot_{\bmf}\vms^2 - 3\CST\vms\dot{\bmf}.
\end{align}

The terms proportional to $\BGE_{\bg}$ can also be discarded in a very similar way, but for that we have to use the full second order Lagrangian as we shall show in the next section. For the moment, the second order matter Lagrangian becomes
\begin{equation} \label{eq:redef:dLAm2}
\begin{split}
\dLA_\mat^{(2)} &= \dLA_{\mat,\text{eff}}^{(2)} + \sqrt{-\bg}\Bigg[\frac{\vms^2}{2}\left(\bed + \bp\right)\BGE_{\bg}\\
&-\left(\B_\mu\B^\mu - \phi^2- 2\C\phi\right)\frac{\bed}{2} - \left(2\C_\mu{}^\nu\C^\mu{}_\nu-\C^2\right)\frac{\bp}{2}\Bigg],
\end{split}
\end{equation}
where the effective matter Lagrangian is expressed by
\begin{equation}
\begin{split}
\frac{\dLA_{\mat,\text{eff}}^{(2)}}{\sqrt{-\bg}} &= \frac{\left(\ded-\bEX\dot{\bmf}^2\vms\right)^2}{2} + \dEX\dot{\bmf}^2\vms \\
&+ \frac{\dot{\bmf}^2}{2}\left(\vms\blapK\vms-\frac{3\kp\dot{\bmf}^2\vms^2}{2}\right).
\end{split}
\end{equation}

\section{Full Second Order Action}

In the last two sections we have developed the perturbative expansion of the action up to second order independently for the gravitational and the matter parts. This was advantageous due to the extensive number of variables. However, we shall now combine them to deal with the full second order Lagrangian for the scalar sector in order to conclude the simplification procedure. Grouping $\dLA^{(2)}=\dLA_\g^{(2,s)}+ \dLA_\mat^{(2)}$ coming respectively from equations Eq.~\eqref{eq:FLRW:dLA2:s} and \eqref{eq:redef:dLAm2} we have
\begin{align}\label{eq:def:dLA2_1}
\frac{\dLA^{(2)}}{\sqrt{-\bg}} =& \frac{1}{3\kp}\left[\blap\dSHs\blapK\dSHs - \left(\dEX-\frac{3\kp\dot{\bmf}^2\vms}{2}\right)^2\right] \nonumber\\
&+ \frac{\left(\ded-\bEX\dot{\bmf}^2\vms\right)^2}{2} + \left(\frac{\CS}{2}-\A\right)\frac{\dSR}{2\kp} \nonumber\\
&+ \frac{\dot{\bmf}^2}{2}\vms\blapK\vms +\frac{\BGE_{\bn}}{2\kp}\left(\B_\gamma\B^\gamma - \phi^2 - 2\C\phi\right) \nonumber\\
&+ \frac{\BGE_{\bh}}{2\kp} \left(2\C_\gamma{}^\nu\C_\nu{}^\gamma-\C^2\right)+\frac{\vms^2}{2}\dot{\bmf}^2\BGE_{\bg}.
\end{align}

The last three terms can be discarded with the same kind of transformation we have used to eliminate the $\BGE_{\bmf}$ in Eqs.~\eqref{eq:def:lgmf2}-\eqref{eq:def:lmf2}. This simplification can be implemented by the following change of variables
\begin{equation}\label{eq:ACtrans}
\A \rightarrow \A + \frac{\dot{\bmf}^2}{2}\vms^2, \qquad \C_\mu{}^\nu \rightarrow \C_\mu{}^\nu + \frac{\dot{\bmf}^2}{2}\vms^2\ \bh_\mu{}^\nu.
\end{equation}

Notice that we can remove the terms proportional to the background field equations even if these equations are not valid, i.e. $\BGE_{\bmf}\neq 0$, $\BGE_{\bn}\neq 0$ and $\BGE_{\bh}\neq 0$. In general, any term linearly proportional to the background equations appearing in the second order action can be removed with this kind of transformation. The first order total Lagrangian can be written as
\begin{equation}\label{eq:dLA1}
\frac{2\kp\dLA_\g^{(1)}}{\sqrt{-\bg}} = -\left[\bG^{\mu\nu} - \kp \bT^{\mu\nu}\right]\ddg_{\mu\nu} = -2\A \BGE_\bn - 2\C \BGE_{\bh}.
\end{equation}
Thus, the prescription to remove these terms is as follows. Any term in the second order Lagrangian that is linear in the $\BGE$'s can be written as
$$ \AQT{1}\BGE_{\bn} + Q^{(2)\mu\nu}\bh_{\mu\nu} \BGE_{\bh} + Q^{(3)}\BGE_{\bg} + Q^{(4)}\BGE_{\bmf},$$
with $\AQT{1}$, $\AQT{2}^{\mu\nu}$, $\AQT{3}$ and $\AQT{4}$ are arbitrary second order tensor fields. These terms can be removed by implementing the transformations
\begin{eqnarray}\label{eq:transgeral}
\A &\rightarrow& \A -\frac{\AQT{1}}{2} - \AQT{3},\\
\C_\mu{}^\nu &\rightarrow& \C_\mu{}^\nu -\frac12 \AQT{2}_\mu{}^{\nu}-\frac13 \AQT{3}\bh_\mu{}^\nu ,\\
\vms &\rightarrow& \vms - \AQT{4}.
\end{eqnarray}

Henceforth, any term linear in the background equations appearing in the second order Lagrangian will be discarded by using the appropriate transformation as described above.

Nevertheless, even after removing the background equations, the $\dLA^{(2)}$ given by Eq.~\eqref{eq:def:dLA2_1} is still not in its most compact form. One should note, for instance, that two of the perturbed variables, namely $\A$ and $\BS$, do not appear with time derivatives. Indeed, they play the role of Lagrange multipliers. This fact becomes evident when we perform the Legendre transformation to go to the Hamiltonian formalism. In addition, this transition naturally leads us to define new variables that will simplify the physical description of the system. 

Thus, instead of introducing a non-evident change of variables in the Lagrangian scenario, we shall now follow the procedure to go from a Lagrangian to a Hamiltonian formalism which will result in the simplest and final form of the second order action.

At the present stage, it seems that there are five variables to describe the scalar perturbations $(\A,\BS,\vms,\CST,\CSD)$. However, as just mentioned, the canonical momentum associated with $\A$ and $\BS$ will result in constraint equations. 

To deal with these constraints we shall follow the procedure developed in \cite{Faddeev1988,Jackiw1993}. This formalism is completely equivalent to Dirac's \cite{Dirac1931,Dirac1949} but has the advantage of being less laborious. Their crucial difference is that we shall implement the Legendre transformation only to those variables that we can complete the method without generating constraint equations.

Accordingly, the canonical momenta associated with $\CST$, $\CSD$ and $\vms$ are
\begin{align}\label{eq:def:pCST}
\pCST &= -\frac{2\sqrt{-\bg}}{\kp}\left(\dEX-\frac{3\kp\dot{\bmf}^2}{2}\vms\right),\\ \label{eq:def:pCSD}
\pCSD &= -\frac{2\sqrt{-\bg}\dSH}{3\kp}, \\ \label{eq:def:pvus}
\pvms &= -\sqrt{-\bg}\left(\ded-\bEX\dot{\bmf}^2\vms\right).
\end{align}

These relations can be inverted to give
\begin{align}
\dot{\CST} &= \frac{1}{3}\left(\frac{3\kp\dot{\bmf}^2}{2}\vms - \frac{\kp\pCST}{2\sqrt{-\bg}} - \bEX\phi - \bscd^2\BS\right),\\
\pct\left(\bscd^2\CSD\right) &= \frac{3\kp}{2\sqrt{-\bg}}\bscd^2\pCSD + \bscd^2\BS,\\
\dot{\vms} &= \frac{\pvms}{\sqrt{-\bg}\dot{\bmf}^2} + \phi - \frac{\BGE_{\bmf}}{\dot{\bmf}}\vms.
\end{align}
By performing the Legendre transformation only to these variables we obtain
\begin{align*}
\dLA^{(2)} &= \pCST\dot{\CST} + \blapK\pCSD\pct\left(\bscd^2\CSD\right) + \pvms\dot{\vms} - \dHA^{(2)}, \\
\end{align*}
with
\begin{align*}
\dHA^{(2)} =& \frac{\pvms^2}{2\sqrt{-\bg}\dot{\bmf}^2}-\frac{\kp\pCST^2}{12\sqrt{-\bg}} + \frac{3\kp\blap\pCSD\blapK\pCSD}{4\sqrt{-\bg}} \\ 
&-\frac{\sqrt{-\bg}}{2\kp}\frac{\CS}{2}\dSR + \frac{\kp\dot{\bmf}^2}{2}\pCST\vms \nonumber\\
&- \frac{\sqrt{-\bg}\dot{\bmf}^2}{2}\vms\blapK\vms- \frac{\BGE_{\bmf}}{\dot{\bmf}}\pvms\vms \\
&+\phi\left[\pvms - \frac{\bEX\pCST}{3} + \sqrt{-\bg}\frac{\dSR}{2\kp}\right]\nonumber\\
&+\BS\blap\left[\blapK\pCSD - \frac{\pCST}{3}\right].
\end{align*}

Once again, the term in the above second order Hamiltonian that is proportional to $\BGE_{\bmf}$ can be eliminated by a redefinition of $\vms$ similarly to Eq.~\eqref{eq:vmstrans}. In addition, we can now fully recognize that $\A$ and $\BS$ are indeed pure Lagrange multipliers since there is no time derivative of these variables and they appear only linearly in the Lagrangian. Requiring the action to be stationary with respect to these variables impose constraint relations among the remaining canonical variables\footnote{The Laplace-Beltrami operator $\blap$ has a unique inverse in the case of a definite signature metric and if the manifold is compact or if its domain is composed of only rapidly decaying functions that go to zero at infinity, (see \cite{Stewart1990} and \cite{chavel1984}).}
\begin{equation}
\pvms = \frac{\bEX\pCST}{3} - \sqrt{-\bg}\frac{\dSR}{2\kp} = 0, \quad  \pCST =3 \blapK\pCSD.
\end{equation}
These relations show us that there is only one independent canonical momentum that could for instance be taken as $\pCSD$. However, combining the above constraints we see that $$\pvms = \frac{2\sqrt{-\bg}}{\kp}\ \blapK\left(\CS+\frac{\bEX\kp}{2\sqrt{-\bg}}\pCSD\right),$$ which suggests us to define the following momentum
\begin{equation}\label{eq:def:pi}
\CSI \equiv \CS + \frac{\bEX\kp}{2\sqrt{-\bg}}\pCSD = \CS + \frac{\bEX}{3}\dSHs.
\end{equation}
Note that this quantity is exactly one of the gauge invariant variables originally introduced by Bardeen~\cite{Bardeen1980} and commonly used in the gauge invariant approach. In terms of this new momentum and using the above constraints we have 
\begin{align*}
\pCST\dot{\CST} &+ \blapK\pCSD\pct\left(\blap\CSD\right) + \pvms\dot{\vms} \nonumber\\
=& \frac{6\sqrt{-\bg}}{\kp\bEX}\blapK\CSI\dot{\CSvms} - \frac{\sqrt{-\bg}}{4\kp}\left[1-\frac{3\dot{\bEX}}{\bEX^2}\right]\CS\dSR \nonumber\\
&- \frac{2\sqrt{-\bg}}{\kp\bEX}\blapK\CSI\dot{\bEX}\vms,
\end{align*}
plus a surface term that we have neglected, where $\CSvms \equiv \CS + {\bEX\vms}/{3}$. We can identify in the above expression that the generalized velocity associated with the canonical momentum $\CSI$ is not just $\dot{\CS}$ nor $\dot{\vms}$ but the variable $\CSvms$. In terms of these new variables the second order Lagrangian reads
\begin{align*}
\dLA^{(2)} &= \frac{6\sqrt{-\bg}}{\kp\bEX}\blapK\CSI\ \dot{\CSvms} - \dHA^{(2)}, \\
\end{align*}
where now
\begin{equation*}
\begin{split}
\frac{\kp\dHA^{(2)}}{\sqrt{-\bg}} =& \frac{2\blap\CSI\blapK\CSI}{\kp\dot{\bmf}^2} + \frac{6\bK}{\kp\dot{\bmf}^2}\left(\frac{\dot{\bEX}}{\bEX^2}+1\right)\CSI\blapK\CSI \\
&- \frac{9\kp\dot{\bmf}^2}{2\bEX^2} \left(\CSvms-\frac{2\bK}{\kp\dot{\bmf}^2}\CSI\right) \blapK \left(\CSvms-\frac{2\bK}{\kp\dot{\bmf}^2}\CSI\right) \\
&+ \BGE_{\bg}\left(\frac{3\CS\blapK\CS}{\bEX^2} + \frac{2\vms\blapK\CSI}{\bEX}-\frac{6\bK}{\bEX^2\kp\dot{\bmf}^2}\CSI\blapK\CSI\right).
\end{split}
\end{equation*}

Once again, the last term being proportional to the background equations $\BGE_{\bg}$ can be discarded with a change of variables. We also note that the variable $\CSvms$ appears only in a specific combination. Thus, we can define a new perturbed variable
\begin{equation}
\MS \equiv \CSvms-\gamma_1\CSI, \qquad \gamma_1 \equiv \frac{2\bK}{\kp\dot{\bmf}^2}.
\end{equation}

As a final simplification, we re-write the term proportional to $\CSI\blapK\CSI$ as
\begin{align*}
\frac{6\bK}{\kp\dot{\bmf}^2}&\left(\frac{\dot{\bEX}}{\bEX^2}+1\right)\CSI\blapK\CSI \\
&= -\frac{3\pct}{\sqrt{-\bg}}\left[\frac{\sqrt{-\bg}}{\bEX}\gamma_1\CSI\blapK\CSI\right] \\
&+ \frac{6\CSI\blapK\pct(\gamma_1\CSI)}{\kp\bEX} - 6 \left(\frac{\pot_{\bmf}}{\dot{\bmf}}-\frac{\BGE_{\bmf}}{\dot{\bmf}}\right)\gamma_1\frac{\CSI\blapK\CSI}{\bEX}.
\end{align*}

This term modifies the generalized velocity associated with $\CSI$ by including precisely the term to transform $\dot{\CSvms}$ into $\dot{\MS}$. Therefore, after neglecting another surface term and performing a final change of variables to remove the term proportional to $\BGE_{\bmf}$ we have that the second order Lagrangian reads
\begin{equation}\label{eq:dLA2}
\dLA^{(2)} = \frac{6\sqrt{-\bg}}{\kp\bEX}\blapK\CSI\ \dot{\MS} - \dHA^{(2)},
\end{equation}
with the unconstrained Hamiltonian $\dHA^{(2)}$ given by
\begin{align}\label{eq:dHA2}
\dHA^{(2)} &= \frac{\sqrt{-\bg}}{\kp}\left[\frac{2\blap_\gamma\CSI\blapK\CSI}{\kp\dot{\bmf}^2} - \frac{9\kp\dot{\bmf}^2}{2\bEX^2}\MS\blapK\MS\right], \\
\blap_\gamma &\equiv \bscd^2 - \gamma_2, \quad \gamma_2 = \frac{6\bK\pot_{\bmf}}{\dot{\bmf}\bEX}.
\end{align}

The above result completes our goal. We have succeeded in obtaining the simplest form of the second order action without ever using the background field equations. As expected, the scalar sector of the second order action has only a single degree of freedom that combines in a specific manner the gravitational and the matter perturbations.

A careful reader might have noticed that while constructing the second order Hamiltonian \eqref{eq:dHA2} we have performed the Legendre transformations only in the perturbed variables. 

In usual perturbative schemes, one normally use the background fields equations, hence, the first order Lagrangian vanishes automatically. However, inasmuch we have nowhere assumed the validity of the background equations of motion, this part still remains. Notwithstanding, we can also avoid this kind of complication by a simple redefinition of the background variables.

Once we have arrived at the Lagrangian given by Eq.~\eqref{eq:dLA2}, we can proceed to construct the zero order Hamiltonian by defining
\begin{equation*}
\pa = \frac{\partial \LA^{(0)}}{\partial \dot{a}},\qquad \pbmf = \frac{\partial \LA^{(0)}}{\partial \dot{\bmf}}.
\end{equation*}

In general, in addition to the zero and second order terms, the full action still contains first order terms. There are different approach to deal with such terms as we have discussed in \cite{Vitenti2012a}. In this work we let this issue aside as our main objective is to build the second order terms without enforcing the classical background motion equations. Hence, ignoring the first order terms, the final result is 
\begin{equation*}
\begin{split}
\LA &= \pa\dot{a} + \pbmf\dot{\bmf} \\
&+ \frac{6\sqrt{-\bg}}{\kp\bEX}\int\dd^3x\blapK\CSI\dot{\MS} - \HA^{(0)} - \int\dd^3x \dHA^{(2)},
\end{split}
\end{equation*}
with 
\begin{equation}
\label{zeroth}
\HA^{(0)} = \frac{\pbmf^2}{2a^3\cV} - \frac{\kp\pa^2}{12Va} - \frac{3\cV a\cK}{\kp},
\end{equation}
where $\cV$ is the comoving volume and $\cK \equiv a^2\bK$ the comoving spatial curvature. The Poisson structure of this system is given by the expressions 
\begin{align*}
\Poisson{a}{\pa} &= 1, &\Poisson{\bmf}{\pbmf} = 1, \\
\Poisson{\MS(x)}{\pMS(y)} &= \delta^3\left(x-y\right), \quad &\mbox{any other is zero .}
\end{align*}

The Lagrangian given by Eq.~\eqref{eq:dLA2} and Eq.~\eqref{eq:dHA2} is first order in time derivatives. If desired, one can obtain a second order time derivative Lagrangian by recovering the relation between $\dot{\MS}$ and $\CSI$, hence, by varying the action with respect to the latter, i.e.,
\begin{equation}
\CSI = \frac{3\kp\dot{\bmf}^2}{2\bEX}\bscd^{-2}_\gamma\dot{\MS}.
\end{equation}

Substituting the above result back in Eq.~\eqref{eq:dLA2} we have
\begin{equation}
\dLA^{(2)} = \frac{1}{2\kp}\left[\frac{\dot{\MS}\blapK\bscd^{-2}_\gamma\dot{\MS}}{\zz^2} + \frac{\MS\blapK\MS}{\zz^2}\right],
\end{equation}
where we have defined
\begin{equation}
\zz^2 \equiv \frac{2\bEX^2}{9\sqrt{-\bg}\kp\dot{\bmf}^2}.
\end{equation}

Our results agree with that obtained in \cite{Garriga1999}, with the difference that in this reference they have included the spatial differential operators $\blapK$ and $\bscd^2_\gamma$ in the definition of $\zz^2$. The procedure of recasting the Lagrangian in the first order form and then solving the constraints was developed in~\textcite{Faddeev1988} (see also~\cite{Jackiw1993}). This was already done in the context of perturbations around a FLRW universe in \cite{Garriga1998} (see Appendix B.1).

As said above, in our calculation we have nowhere used the background field equations to simplify the Lagrangian. All simplifications have been done by identifying terms proportional to the background equations and redefining the perturbed variables. Note that this is not equivalent to impose the validity of the background equations. The only type of terms that we can eliminate following this reasoning are those linearly proportional to the background field equations. Any term having, for instance, time derivative of perturbations cannot be discarded with our procedure.

\section{Conclusions}

Without using any background Einstein's equation, we were able to obtain the very simple second order Hamiltonian Eq.~\eqref{eq:dHA2} describing the dynamics of linear cosmological perturbations on homogeneous and isotropic geometries with generic spacelike hypersurfaces for the case of a canonical scalar field with an arbitrary potential. This is a generalization of previous works. The resulting Hamiltonian essentially coincides with the one obtained in the literature if one assumes the validity of the background Einstein's equations. 

This second order Hamiltonian, together with its zeroth order companion presented Eq.~\eqref{zeroth}, can now be used to investigate the evolution of scalar perturbations of general canonical scalar fields in the situation where the background is also quantized, an improvement over the usual semi-classical approach to the inflationary scenario. Then we can move to more involved subjects, like obtaining the Hamiltonian describing the dynamics of linear perturbations in Bianchi models and of second order perturbations in Friedmann models. These will be the subjects of our future investigations.

\section*{ACKNOWLEDGEMENTS}
We would like to thank CNPq of Brazil for financial support.  We also would like to thank `Pequeno Seminario' of CBPF's Cosmology Group for useful discussions, comments and suggestions.
\bibliographystyle{apsrev}

\end{document}